\documentclass[aps,prl,epsfig,floats,twocolumn,amssymb,amsmath,floatfix,showpacs]{revtex4}
\usepackage{graphicx}
\usepackage{bm}  
\usepackage{amssymb}
\usepackage{color}
\usepackage{amscd}

\begin{document}

\title{Anomalous Diffusion of Symmetric and Asymmetric Active Colloids}

\author{Ramin Golestanian}
\email{r.golestanian@sheffield.ac.uk} \affiliation{Department of
Physics and Astronomy, University of Sheffield, Sheffield S3 7RH,
UK}

\date{\today}

\begin{abstract}
The stochastic dynamics of colloidal particles with surface
activity---in the form of catalytic reaction or particle
release---and self-phoretic effects is studied analytically. Three
different time scales corresponding to inertial effects, solute
redistribution, and rotational diffusion are identified and shown to
lead to a plethora of different regimes involving inertial,
propulsive, anomalous, and diffusive behaviors. For symmetric active
colloids, a regime is found where the mean-squared displacement has
a super-diffusive $t^{3/2}$ behavior. At the longest time scales, an
effective diffusion coefficient is found which has a non-monotonic
dependence on the size of the colloid.
\end{abstract}

\maketitle

The development of biomimetic technology would significantly benefit
from the ability to make synthetic components with desired motility
properties. In recent years, there has been a range of developments
along these lines, with functionalities that can be manipulated at
length scales ranging from microns down to molecular scales
\cite{leigh-etal}. These include experimental realization of
actuated microswimmers \cite{Dreyfus} and theoretical proposals of
simple model swimmers that can tackle the problems caused by low
Reynolds number conditions \cite{swimmer}. Moreover, it has been
recently demonstrated (both experimentally and theoretically) that
interfacial phoretic effects (such as electrophoresis,
electroosmosis, and diffusiophoresis) could lead to self-propulsion
of colloidal particles \cite{phoretic,jon,gla}. It has also been
shown that phoretic effects can be used to steer both active and
passive colloidal particle \cite{chemo,lyderic}, which adds to the
promise of these phenomena for designing functional self-motile
vessels in the nanoscale.

A fundamental property of such small objects, even when equipped
with a self-propulsion mechanism, is that their motion is stochastic
due to the ambient fluctuations that could be of thermal origin or
otherwise. This means that we cannot directly control the motion of
self-motile objects, and any design characteristic needs to be
incorporated into statistical average outcomes. For example, it has
been shown that a self-propelled colloidal particle makes a
crossover between ballistic and diffusive behaviors over a time
scale that is set by the rotational diffusion of the colloid, when
its orientation will be randomized \cite{jon}. A study of a two
dimensional model of self-propelled objects with fluctuations both
in direction and magnitude of the velocity has shown the possibility
of re-entrant ballistic and diffusive behaviors \cite{2D-prl}. Other
examples include the effect of the activity of proteins on the
dynamics of membranes \cite{act-mem}, and collective behavior of
mixtures of motors and filaments (or reorganizing living cells) and
active particles such as swimming bacteria
\cite{active-medium-th,upenn}, where a host of qualitative and
quantitative changes have been found to occur due to nonequilibrium
fluctuations. In light of this inherent feature, it will be natural
to ask how many distinct regimes of motion we could have for active
colloidal particles, what the relevant time scales that
differentiate between these regimes are, and how they can be tuned
so that the desired type of motion could be achieved by choosing the
right parameters.

\begin{figure}[b]
\includegraphics[width=.76\columnwidth]{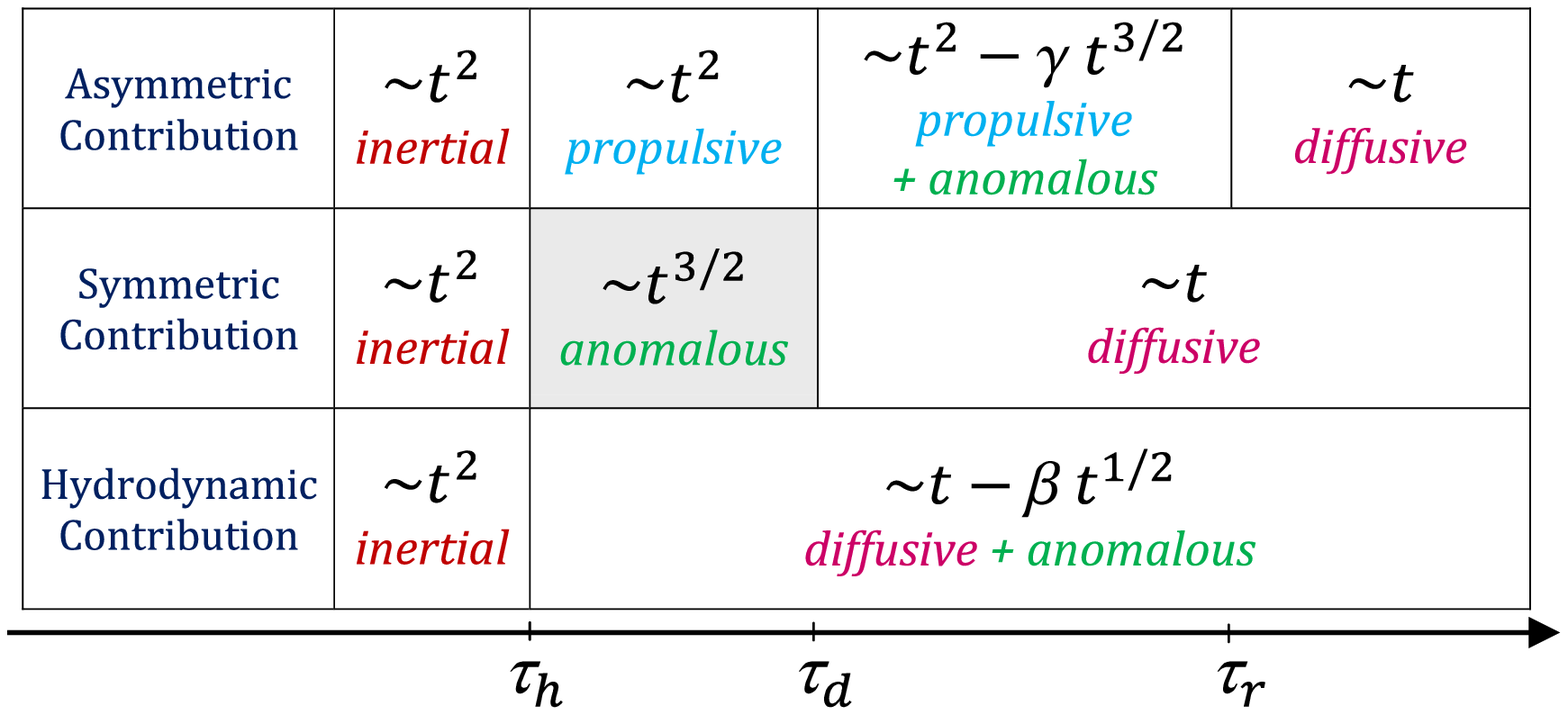}
\caption{Summary of results for the different contributions to the
mean-squared displacement of active colloids. The total mean-squared
displacement is obtained by summing all of the contributions for
asymmetric colloids, and the bottom two rows (only) for symmetric
colloids.} \label{fig:summ}
\end{figure}

Here, we aim to address some of these questions for a class of
isolated self-motile active colloids. We study the velocity
autocorrelation function and the mean-squared displacement of
surface-active spherical colloidal particles that interact with
their self-generated surrounding clouds of solute particles via
interfacial phoretic effects. We identify the relevant time scales
in the dynamics, namely the hydrodynamic relaxation time $\tau_h$
that controls the crossover between inertial and viscous regimes,
the diffusion time of the solute particles around the colloid
$\tau_d$, and the rotational diffusion time of the colloid $\tau_r$.
We calculate the contribution due to hydrodynamic fluctuations, as
well as the self-phoretic contributions that depend on whether the
particles are symmetric (in which case there is no net propulsion)
or asymmetric (where the colloids are self-propelled). We find that
these different contributions lead to a variety of different
regimes, as summarized in Fig. \ref{fig:summ}. For symmetric surface
activity, we find a regime corresponding to $\tau_h \ll t \ll
\tau_d$ where the active colloid demonstrates a super-diffusive
behavior with a mean-squared displacement $\sim t^{3/2}$ (shaded
region in Fig. \ref{fig:summ}). This is similar and somewhat related
to an anomalous super-diffusive regime found in active bacterial
suspensions \cite{upenn}. Other regimes include inertial $t^2$ or
propulsive $t^2$ behaviors, diffusive behavior $\sim t$, and an
anomalous correction of the form $-\gamma t^{3/2}$.

\begin{figure}
\includegraphics[width=.85\columnwidth]{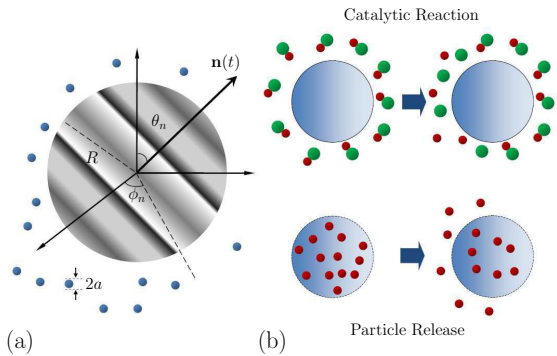}
\caption{(a) Schematics of an instantaneous configuration of an
axially symmetric surface-active spherical colloid. (b) The model
could correspond to a chemical reaction catalyzed on the surface
with the simplifying assumption that one of the product particles is
very similar to the substrate, or a container that releases
particles through channels.} \label{fig:schem}
\end{figure}

We consider a spherical colloidal particle of radius $R$ with an
axially symmetric pattern of surface activity as shown in Fig.
\ref{fig:schem}a, which in essence leads to the release of (excess)
product particles ${\cal P}$ with diffusion coefficient $D$. This
could correspond to a chemical reaction ${\cal S} \longrightarrow
{\cal S}'+{\cal P}$ catalyzed on the surface of the colloid, with
the simplifying assumption that the produced ${\cal S}'$ will act
almost like the consumed ${\cal S}$ (see Fig. \ref{fig:schem}b). On
the other hand, it could also correspond to a system that actively
releases ${\cal P}$ particles from the interior of the sphere.
The product particles will have a concentration profile $C({\bf
r},t)$ that could interact with the fluid in the vicinity of the
colloid due to surface phoretic effects, and cause relative motion
with a slip velocity ${\bf v}_s(\theta,\phi,t)=\mu
\nabla_{\parallel} C(R,\theta,\phi,t)$ (in standard spherical
coordinates; see Fig. \ref{fig:schem}a), where $\mu$ is the surface
mobility and $\nabla_{\parallel}$ denote the lateral gradient
\cite{Anderson-review}. In diffusiophoresis, $\mu=k_{\rm B} T
\lambda^2/\eta$, where $k_{\rm B} T$ is the thermal energy scale,
$\eta$ is the viscosity of water, and $\lambda$ is the Derjaguin
length \cite{Derjaguin}, which is defined in terms of the
interaction potential $W(z)$ between the diffusing particles and the
surface of the colloid as
\begin{math}
\lambda^2=\int_{0}^\infty d z \;z \left[1-{\rm e}^{-{W(z)}/{k_{\rm
B} T}}\right].
\end{math}
Note that in our notations, $\lambda^2$ will be positive (negative)
for repulsive (attractive) effective surface potentials. In the case
of ionic systems $\lambda$ will be set by the Debye length
\cite{lyderic,Anderson-review}. Averaging over the surface of the
sphere (using integration over the solid angle $\Omega$), we can
obtain the instantaneous velocity of the colloid as ${\bf
v}(t)=-\frac{1}{4 \pi} \int d \Omega \;\mu \nabla_{\parallel}
C(R,\theta,\phi,t)$.

The axis of symmetry of the colloid, which points to the direction
of propulsion for sufficiently asymmetric patterns \cite{gla}, is
defined by the unit vector ${\bf n}(t)=(\sin \theta_n(t) \cos
\phi_n(t),\sin \theta_n(t) \sin \phi_n(t),\cos \theta_n(t))$ (see
\ref{fig:schem}a). The stochastic nature of ${\bf n}(t)$ due to
rotational diffusion of the colloid causes the cloud of product
particles to constantly redistribute, which will in turn make the
velocity of the active colloid fluctuate. To get the instantaneous
velocity of the colloid, we need to solve the diffusion equation for
the concentration profile of the product particles, namely
\begin{equation}
\partial_t C({\bf r},t)-D \nabla^2 C({\bf r},t)
=\alpha(\theta,\phi,t) \delta(r-R),\label{diff-eq-1}
\end{equation}
subject to the boundary condition of vanishing normal current on the
surface of the sphere. In Eq. (\ref{diff-eq-1}),
$\alpha(\theta,\phi,t)$ is the surface activity function of the
sphere, {\em i.e.} rate per unit area of the introduction of
(excess) product particles. For axially symmetric surface activity,
we can represent the function in terms of the spherical harmonics as
\begin{math}
\alpha(\theta,\phi,t)=\sum_{\ell,m} \left(\frac{4 \pi }{2
\ell+1}\right) \alpha_{\ell} \; Y^*_{\ell m}(\theta_n(t),\phi_n(t))
Y_{\ell m}(\theta,\phi).
\end{math}
Equation (\ref{diff-eq-1}) only gives the average density, and the
linear relation between the velocity and the concentration profile
suggests that in order to calculate velocity correlations we need to
incorporate the density fluctuations as well, which we do using the
method outlined in Ref. \cite{gla}.

Using the formulation described above we can calculate the velocity
autocorrelation function for the active colloid
\begin{math}
A^{vv}(t) \equiv \left \langle {\bf v}(t) \cdot {\bf v}(0) \right
\rangle
\end{math}
as well as the mean-squared displacement
\begin{math}
\Delta L^2(t) \equiv \left \langle [{\bf r}(t)-{\bf r}(0)]^2 \right
\rangle=\int_0^t d t_1 \int_0^t d t_2 \left \langle {\bf v}(t_1)
\cdot {\bf v}(t_2) \right \rangle.
\end{math}
There are three important time scales in the problem. The
characteristic diffusion time of the product particles around the
sphere is $\tau_d=R^2/D$, where $D={k_{\rm B} T}/{(6 \pi \eta a)}$
depends on the radius of the solute particles $a$. This time scale
sets the relaxation time of the redistribution of the particles
around the sphere when it changes orientation. The rotational
diffusion time, $\tau_r=4 \pi \eta R^3/k_{\rm B} T$, controls the
changes in the orientation of the sphere, and is defined via the
orientation autocorrelation function: $\langle {\bf n}(t) \cdot {\bf
n}(0) \rangle=e^{-t/\tau_r}$ \cite{note}. Finally, the hydrodynamic
time that controls the crossover between the inertial and the
viscous regimes is given as $\tau_h=R^2/\nu$, where $\nu=\eta/\rho$
is the kinematic viscosity of water that involves the mass density
$\rho$. Practically speaking, we always have $\tau_h \ll \tau_d \ll
\tau_r$ although this is not a fundamental requirement.

We can identify three distinct contributions to the velocity
autocorrelation function: (1) a contribution from the density
fluctuations, which turns out to be only sensitive to the overall
symmetric component of the activity and is present even for
non-propelled active colloids, (2) a contribution from the
asymmetric component of the activity, and (3) a hydrodynamic
contribution that entails the passive diffusion of the colloid and
the hydrodynamic long-time tail. These will lead to distinct
contributions to the mean-squared displacement, which will add up to
make the total mean-squared displacement, namely
\begin{math}
\Delta L^2(t)=\Delta L_{\rm sym}^2(t)+\Delta L_{\rm
asym}^2(t)+\Delta L_{\rm hyd}^2(t).
\end{math}
We will focus on each of these contributions separately below.

\paragraph{Symmetric Contribution.---}

The density fluctuations that are accounted for by adding a noise
term to Eq. (\ref{diff-eq-1}) lead to a contribution to the velocity
autocorrelation function that is proportional to $\alpha_0$
\cite{ramin}---the $\ell=0$ coefficient in the expression for
$\alpha(\theta,\phi,t)$ in terms of the spherical harmonics
[$\alpha_0=\frac{1}{4 \pi} \int d \Omega \;\alpha(\theta,\phi,t)$].
This means that the contribution by density fluctuations is only
sensitive to the mean overall surface activity of the colloid, and
not the patterning structures on it. We find the asymptotic
behaviors $A^{vv}_{\rm sym} \simeq \frac{\alpha_0 \mu^2}{\pi^{3/2} D
R^4} (t/\tau_d)^{-1/2}$ for $t \ll \tau_d$ and $A^{vv}_{\rm sym}
\simeq \frac{3 \alpha_0 \mu^2}{32 \pi^{3/2} D R^4}
(t/\tau_d)^{-5/2}$ for $t \gg \tau_d$. We also calculate the
contribution of density fluctuations to the mean-squared
displacement, which has the asymptotic behaviors of
\begin{equation}
\Delta L_{\rm sym}^2 \simeq \frac{8 \alpha_0 \mu^2}{3 \pi^{3/2}
D^{3/2} R^3} \; t^{3/2}\;\;\; ; \; t \ll \tau_d,\label{MSD-sym-1}
\end{equation}
at short times, and
\begin{equation}
\Delta L_{\rm sym}^2 \simeq \frac{2 c_1 \alpha_0 \mu^2}{\pi^{2}
D^{2} R^2} \; t \;\;\; ; \; t \gg \tau_d,\label{MSD-sym-2}
\end{equation}
at long times, and a smooth crossover between them. Here,
$c_1=1.17810$ is a numerical prefactor.

\paragraph{Asymmetric Contribution.---}

Solving Eq. (\ref{diff-eq-1}) without the noise term, we can
calculate the propulsion velocity of the colloid as a function of
time for a given time dependent orientation trajectory. We find
\begin{math}
{\bf v}(t)=\frac{v_0}{\tau_d} \int_{-\infty}^t d t' {\cal M}(t-t')
{\bf n}(t'),\label{vMn-def}
\end{math}
where $v_0=-\alpha_1 \mu/(3 D)$ is the mean propulsion velocity
\cite{gla}, and the {\em memory} kernel is given as
\begin{math}
{\cal M}(t)=\frac{2}{\pi} \int_0^\infty d u \frac{u^{3/2}}{(u^2+4)}
e^{-u (t/\tau_d)},
\end{math}
with asymptotic behaviors ${\cal M}(t) \simeq \frac{2}{\sqrt{\pi}}
(t/\tau_d)^{-1/2}$ for $t \ll \tau_d$ and ${\cal M}(t) \simeq
\frac{3}{8 \sqrt{\pi}} (t/\tau_d)^{-5/2}$ for $t \gg \tau_d$. Note
that the propulsion velocity is controlled by the $\ell=1$ term
($\alpha_1$) in the surface activity profile.

Rotational diffusion of the colloid randomizes its orientation over
the time scale $\tau_r$, which leads to a contribution to the
velocity autocorrelation function of the form of a convolution
between two memory kernels and the orientation autocorrelation
function. This leads to a velocity autocorrelation function
\cite{ramin}, which have three different regimes, due to the
presence of two characteristic time scales $\tau_d$ and $\tau_r$. At
short times, $t \ll \tau_d \ll \tau_r$, we find $A^{vv}_{\rm asym}
\simeq v_0^2 \left[1-\frac{4 c_2 }{\pi}
\frac{\tau_d}{\tau_r}-\frac{1}{2} \frac{\tau_d^{3/2}}{\tau_r^{5/2}}
t+\frac{4}{\pi} \frac{t^2}{\tau_d \tau_r}
\ln\left(\frac{t}{\tau_d}\right)\right]$ where $c_2=0.642699$ is a
numerical prefactor, which implies that the autocorrelation function
will be rounded off at small $t$. For intermediate time, $\tau_d \ll
t \ll \tau_r$, we find $A^{vv}_{\rm asym} \simeq v_0^2
\left[1-\frac{t}{\tau_r}-\frac{1}{\sqrt{\pi}}
\frac{\tau_d^{3/2}}{\tau_r} t^{-1/2}\right]$, and for long times,
$\tau_d \ll \tau_r \ll t$, we find $A^{vv}_{\rm asym} \simeq v_0^2
\left[e^{-t/\tau_r}+\frac{3}{4 \sqrt{\pi}} \tau_r \tau_d^{3/2}
t^{-5/2}\right]$, which means that the decay at long times is
primarily algebraic and not exponential. Consequently, the
mean-squared displacement will have three different regimes. We find
the asymptotic form of
\begin{equation}
\Delta L_{\rm asym}^2 \simeq v_0^2 t^2 \left[1-\frac{4
c_2}{\pi}\left(\frac{\tau_d}{\tau_r}\right)\right]\;\;\; ; \; t \ll
\tau_d \ll \tau_r,\label{MSD-asym-1}
\end{equation}
at short times,
\begin{equation}
\Delta L_{\rm asym}^2 \simeq v_0^2 t^2-\left(\frac{8}{3
\sqrt{\pi}}\right)
 \frac{v_0^2 \tau_d^{3/2}}{\tau_r} \; t^{3/2} \;\;\; ; \;  \tau_d \ll
t \ll \tau_r,\label{MSD-asym-2}
\end{equation}
at intermediate times, and
\begin{equation}
\Delta L_{\rm asym}^2 \simeq 2 v_0^2 \tau_r \; t \;\;\; ; \; \tau_d
\ll \tau_r \ll t,\label{MSD-asym-3}
\end{equation}
at long times, with a smooth crossover between them.

\paragraph{Hydrodynamic Contribution.---}

Thermal fluctuations of the solvent fluid velocity also contribute
to the velocity autocorrelation function of the sphere, because of
the no-slip boundary condition between the fluid and the colloid.
Performing a similar calculation, one finds \cite{zwanzig}
\begin{math}
\Delta L_{\rm hyd}^2 \simeq 6 D_0 t-\frac{2 k_{\rm B} T
\rho^{1/2}}{\pi^{3/2} \eta^{3/2}} \; t^{1/2}
\end{math}
for $t \gg \tau_h$, where $D_0={k_{\rm B} T}/{(6 \pi \eta R)}$ is
the bare diffusion coefficient of the colloid. The first term in the
above equation describes the standard passive diffusion of the
sphere while the second term corresponds to the hydrodynamic
long-time tail \cite{alder,schmidt}. At short times when $t \ll
\tau_h$, one finds
\begin{math}
\Delta L_{\rm hyd}^2 \simeq 3 \left(\frac{k_{\rm B} T}{M_{\rm
eff}}\right)t^2
\end{math}
where $M_{\rm eff}$ is the effective inertial mass of the colloid in
water. The above results are summarized in Fig. \ref{fig:summ}.

\paragraph{Discussion.---}

At the longest time scales ($t > \tau_r$), all of the contributions
are diffusive, leading to a total effective diffusion coefficient
\begin{equation}
D_{\rm eff}=\frac{k_{\rm B} T}{6 \pi \eta R}+\frac{4 \pi \alpha_1^2
\mu^2 \eta R^3}{27 D^2 k_{\rm B} T}+\frac{c_1 \alpha_0 \mu^2}{3
\pi^2 D^2 R^2}.\label{Deff}
\end{equation}
The different terms in the above expression exhibit different
$R$-dependencies, which causes the asymmetric contribution to be
dominant for $R \gtrsim \left[{D k_{\rm B} T}/{(\alpha_1 \mu
\eta)}\right]^{1/2}$, while the symmetric contribution takes over
when $R \lesssim {\alpha_0 \mu^2 \eta^2}/({D^2 k_{\rm B} T})$. At
the shortest time scales, on the other hand, the contribution due to
phoretic effects will also be dominated by inertial effects that
should lead to ballistic contributions (see Fig. \ref{fig:summ}).

In the intermediate times, we observe a number of anomalous
behaviors. For symmetric active colloids, the super-diffusive
$t^{3/2}$ behavior [Eq. (\ref{MSD-sym-1})] for  $\tau_h < t <
\tau_d$ is a new regime (for isolated self-propelled particles)
where the motion is neither ballistic nor diffusive. The reason a
symmetric particle can move at all is because density fluctuations
of the cloud of solute particles can instantaneously produce an
asymmetric distribution and therefore a net propulsion in some
direction. This motion, however, will be decorrelated via density
fluctuations themselves, leading to fluctuations without symmetry
breaking. We can understand the form of Eq. (\ref{MSD-sym-1}) as
follows: using $\Delta L^2 \sim v^2 t^2$, and putting $v \sim \mu
\nabla C \sim \mu \delta C/R$, we find $\Delta L^2 \sim \mu^2
\langle \delta C(t) \delta C(0) \rangle t^2/R^2$. The density
auto-correlation function can be written as $\langle \delta C(t)
\delta C(0) \rangle=\langle \delta C^2 \rangle k(t)$, involving the
density fluctuations $\langle \delta C^2 \rangle$ and the kernel
$k(t)$ that controls the relevant relaxation mode \cite{doi}. Here,
relaxation is controlled by diffusion, hence $k(t) \sim 1/(D
t)^{3/2}$, and the number fluctuations are controlled by the average
number of particles ($\langle \delta N^2 \rangle \sim N_{\rm
ave}$)---as inherent to any Poisson process---that yield $\langle
\delta C^2 \rangle \sim C_{\rm ave}$. On the other hand, the average
density is controlled by the average particle production rate (per
unit area) $\alpha_0$ as $C_{\rm ave}\sim (\alpha_0 R^2 t)/R^3$.
Putting these all together, we find Eq. (\ref{MSD-sym-1}). This
shows that the active velocity fluctuations are controlled by two
mechanisms: particle production (that controls the density
fluctuations) and diffusion of the produced particles.
Interestingly, a similar $t^{3/2}$ power law has been observed in
the motion of passive tracer particles in a bath of bacteria whose
flagella stir up the fluid (and theoretically accounted for using a
phenomenological continuum active medium theory) \cite{upenn}, and
in ion channel gating \cite{hanggi}, both of which cases are also
governed by some sort of density fluctuations \cite{note2}.

For asymmetric particles when $\tau_h < t < \tau_d$, the $t^{3/2}$
contribution is added (with a positive coefficient) to the $t^{2}$
propulsive term. On the other hand, for  $\tau_d < t < \tau_r$ the
memory effect that exists for self-propelled asymmetric colloids
introduces an anomalous {\em anti-correlation} (i.e. contribution
with negative sign) in the velocity autocorrelation function and the
mean-squared displacement [Eq. (\ref{MSD-asym-2})]. Such anomalous
corrections, which have also been observed in continuum theories of
interacting active self-propelled particles \cite{active-medium-th},
are reminiscent of the effect of the hydrodynamic long-time tail.
Note, however, that the anomalous $-\gamma t^{3/2}$ correction in
Eq. (\ref{MSD-asym-2}) corresponds to much longer time scales and
should be more easily observable than the hydrodynamic long-time
tail.

To get a better feel for the working domain of each regime, we can
write the time scales (for water at room temperature and using a
typical value of $a=1 ~\AA$) in the following convenient form:
$\tau_h=10^{-6} ~(R/1 \mu{\rm m})^2$ s, $\tau_d=10^{-3} ~(R/1
\mu{\rm m})^2$ s, and $\tau_r=3 ~(R/1 \mu{\rm m})^3$ s. This shows
that while rapid cameras or scattering techniques
\cite{upenn,schmidt} could in principle resolve all the three
domains for micron-sized beads, using $R=20 ~\mu{\rm m}$ (which
yields $\tau_d=0.4$ s and $\tau_r=2.4 \times 10^{4}$ s) should
provide a comfortable working range for an experiment that aims to
resolve the anomalous components of the motion. Finally, we note
that here we have only focused on the phoretic contributions to
velocity fluctuations, and in practice other sources of fluctuations
might also be present \cite{act-mem}, which need to be taken into
account.

In summary, we have shown that active colloidal particles that
interact with their self-generated cloud of solute particles can
have a range of different types of stochastic motions.
Using parameters such as surface activity, surface mobility, and
size, we can tune the behavior of active colloidal particles, and
this could provide new possibilities in designing functional motile
agents for applications in micro- and nano-fluidics and targeted
delivery.

\acknowledgements

I acknowledge invaluable discussions with A. Ajdari, R.A.L. Jones,
and T.B. Liverpool, and support from the EPSRC.


\begin{thebibliography}{99}
\bibitem{note}
Note that there is a factor of two diffrenece in the definition of
the rotational diffusional diffusion time $\tau_r$ here and $\tau_R$
in Ref. \cite{jon}, namely $\tau_r=\tau_R/2$.
\bibitem{note2}
The result presented in Ref. \cite{hanggi} for the residence time
distribution of closed state of the ion channel $f_c(t) \sim
t^{-3/2}$ is the equivalent of the derivative of the velocity
autocorrelation function $A^{vv}(t) \sim t^{-1/2}$, which leads to
$\Delta L^2 \sim t^{3/2}$ law.
\end{thebibliography}
\end{document}